\documentclass[final,5p,times,twocolumn]{elsarticle}

\usepackage{amssymb}
\journal{Nuclear Instruments and Methods A}

\begin{document}

\begin{frontmatter}



\title{High-speed X-ray imaging spectroscopy system 
 with Zynq SoC for solar observations}


\author[isas]{Shin-nosuke Ishikawa}
\author[isas]{Tadayuki Takahashi}
\author[isas]{Shin Watanabe}
\author[naoj]{Noriyuki Narukage}
\author[naoj]{Satoshi Miyazaki}
\author[oist]{Tadashi Orita}
\author[oist]{Shin'ichiro Takeda}
\author[handai]{Masaharu Nomachi}
\author[shimafuji]{Iwao Fujishiro}
\author[shimafuji]{Fumio Hodoshima}

\address[isas]{Institute of Space and Astronautical Science, Japan Aerospace Exploration Agency, 3-1-1 Yoshinodai, Chuo-ku, Sagamihara, Kanagawa 252-5210, Japan}
\address[naoj]{National Astronomical Observatory of Japan, 2-21-1 Osawa, Mitaka, Tokyo 181-8588, Japan}
\address[oist]{Advanced Medical Instrumentation Unit, Okinawa Institute
of Science and Technology Graduate University, 1919-1 Tancha, Onna-son, Kunigami-gun, Okinawa 904-0495, Japan}
\address[handai]{Research center for Nuclear Physics, Osaka-University, 1-1 Machikaneyama, Toyonaka, Osaka 560-0043, Japan}
\address[shimafuji]{Shimafuji Electric Inc., NS building 3F, 6-36-11 Nishi-Kamata, Ota-ku, Tokyo 144-0051, Japan}

\begin{abstract}
We have developed a system combining a back-illuminated Complementary-Metal-Oxide-Semiconductor (CMOS) imaging sensor and Xilinx Zynq System-on-Chip (SoC) device for a soft X-ray (0.5--10~keV) imaging spectroscopy observation of the Sun to investigate the dynamics of the solar corona. Because typical timescales of energy release phenomena in the corona span a few minutes at most, we aim to obtain the corresponding energy spectra and derive the physical parameters, i.e., temperature and emission measure, every few tens of seconds or less for future solar X-ray observations. An X-ray photon-counting technique, with a frame rate of a few hundred frames per second or more, can achieve such results. We used the Zynq SoC device to achieve the requirements. Zynq contains an ARM processor core, which is also known as the Processing System (PS) part, and a Programmable Logic (PL) part in a single chip. We use the PL and PS to control the sensor and seamless recording of data to a storage system, respectively. We aim to use the system for the third flight of the Focusing Optics Solar X-ray Imager (FOXSI-3) sounding rocket experiment for the first photon-counting X-ray imaging and spectroscopy of the Sun.
\end{abstract}


\end{frontmatter}


\section{Introduction}
The solar corona is an active and dynamic outer atmosphere of the Sun with an exceedingly high temperature of a few million Kelvins and low density. The solar corona also embodies various types of energy release phenomena including solar flares \cite{aschwanden2004}. The corona emits X-rays, and its observations are important in understanding the physical processes of such phenomena. X-ray imaging spectroscopy is a desirable tool in deriving physical parameters such as temperature, emission measures, and spectral slope from a plasma with non-thermal energy distribution \cite{whitepaper}. However, imaging spectroscopy of the Sun in the soft X-ray energy range of $<$10 keV, which is a typical energy range of the coronal emission, have not been fully developed yet owing to technical difficulties.

Since timescales of the coronal energy release phenomena are as short as a few tens of seconds, we must develop a spectrum in similar timescales to investigate such phenomena by X-ray imaging spectroscopy observations.  High-speed single photon-counting techniques are a possible solution for the measurement of the energy of each X-ray photon \cite{whitepaper,sakao2014}. To achieve this solution, we must measure the energy of more than one thousand X-ray photons in a region of interest every tens of seconds. The Sun is sufficiently bright in the soft X-ray range to generate a spectrum every tens of seconds for every pixel with X-ray focusing optics. However, Charge-Coupled Devices (CCDs) do not have sufficient readout speed to measure spectra in such short time intervals.

Therefore, photon-counting observations of the Sun have not been performed in the soft X-ray range, and only photon-integrated imaging observations have been performed (e.g., X-ray telescope onboard the Hinode satellite \cite{golub2007}). These instruments performed multi-filter observations to estimate the plasma temperature of the corona. However, this method does not have strong capabilities to detect multi-thermal components and non-thermal plasma. Spectroscopic measurements would be an attractive tool to investigate the details of the coronal plasma in future observations.

High-speed photon counting measurements with a few hundreds of frames per second (fps), by a back-illuminated Complementary Metal Oxide Semiconductor (CMOS) imaging sensor, can overcome this limitation. Back-illuminated CMOS sensors can detect X-rays down to sub-keV energies, and can be operated with significantly higher speed than CCDs. In past applications, we were unable to find a good example of a CMOS sensor with low noise and small charge loss to measure the X-ray photon energy accurately. Recently, we have demonstrated that we can achieve solid performance in order to perform high-speed X-ray imaging spectroscopy with some sensors \cite{narukage2017}. We evaluated a back-illuminated CMOS sensor with 2048 $\times$ 2048 pixels and a pixel size of 11 $\mu$m. A good energy measurement capability for X-rays was achieved from this. We plan to perform soft X-ray imaging and spectroscopy observations using this type of CMOS imaging sensors.

X-ray observation can be performed only from high altitudes, such as from satellites, sounding rockets, and balloons, owing to the absorption of X-rays by the Earth's atmosphere.  Therefore, a compact and vacuum-compatible high-speed data acquisition system is required. In this paper, we present a concept and solution for the solar X-ray imaging spectroscopy observations with a high-speed CMOS sensor and data acquisition system using a Xilinx Zynq All-Programmable (AP) System-on-Chip (SoC) device \cite{zynq}.

\section{X-ray imaging spectroscopy with a sounding rocket}
We will perform the first photon counting imaging and spectroscopy in the soft X-ray range (0.5--10 keV) with the third flight of Focusing Optics Solar X-ray Imager (FOXSI) sounding rocket experiment in the summer of 2018 (FOXSI-3). FOXSI is an international collaboration sounding rocket experiment for high-sensitivity X-ray imaging spectroscopy observation of the Sun by the University of California, Berkeley, University of Minnesota, NASA, and ISAS/JAXA. The past two FOXSI observations were successfully performed with a target energy band of hard X-rays above a few keV \cite{glesener2016,krucker2014,ishikawa2014,ishikawa2017}. FOXSI has seven modules of X-ray grazing incidence telescopes; we used semiconductor hard X-ray detectors for past launches \cite{christe2016, ishikawa2016, ishikawa2011}. Since the FOXSI telescope has a focusing capability to the soft X-ray energy range, we will replace one detector with the CMOS sensor as the focal plane detector for the FOXSI-3 to realize the first soft X-ray high-speed imaging spectroscopy of the Sun. We call this soft X-ray observation project as the Photon Energy Imager in soft X-rays (PhoEnIX) as part of FOXSI-3.

The PhoEnIX observation is not only a scientific observation, but also a conceptual and technical demonstration for future observations including satellite applications. Therefore, we wish to record as much image data as possible for technical evaluations. In the case where we read a 200 $\times$ 500 pixels area in the CMOS sensor for the flight (corresponding to a 200 arcsec $\times$ 500 arcsec area on the Sun with the 2-m focal length of the FOXSI optics) with the highest possible frame rate for the sensor with 500-fps speed \cite{narukage2017} and 12-bit data resolution, we require a data readout and recording speed that is $>$72 MB/s. This data rate is considerably faster than the speed of a telemetry stream (2 Mbps), and data storage is necessary. The observational time of the FOXSI flight is approximately 6 min, and $>$25 GB volume is required for the data storage to record all the observational data.

\begin{figure}[ht!]
\begin{center}
\includegraphics[width=6.5cm]{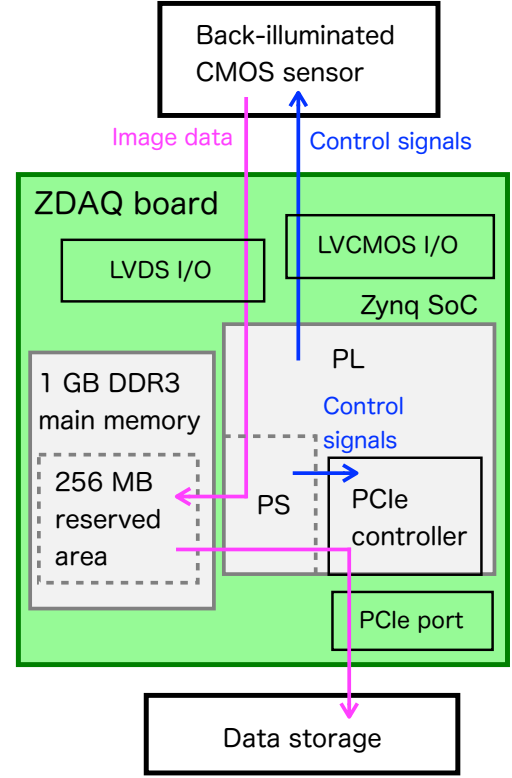}
\end{center}
\caption{Concept diagram of the high-speed X-ray detector system for the sounding rocket experiment.\label{fig:diagram}}
\end{figure}

To control the sensor, synchronous signals are necessary and the Field Programmable Gate Array (FPGA) is a candidate. Nevertheless, it is more reliable to record a large volume of data with high-speed by a processor since the data recording function is widely used and well optimized. To take advantage of the FPGA and processor, we will use a device in the Xilinx Zynq series to achieve the high-speed data acquisition during the flight. Zynq has the flexibility of software (asynchronous) and high-speed of hardware (synchronous) by combining Processing System (PS) and Programmable Logic (PL) parts in a single chip. A close connection between PS and PL provides increased bandwidth. As such, we can achieve the high-speed X-ray imaging and spectroscopy by using PL to control the sensor and receive data and PS to record the data using an operating system. The concept of the CMOS readout system to meet the requirements is illustrated in Fig.~\ref{fig:diagram}.  

\section{Design of the data acquisition board}

\begin{figure}[ht!]
\begin{center}
\includegraphics[width=8cm]{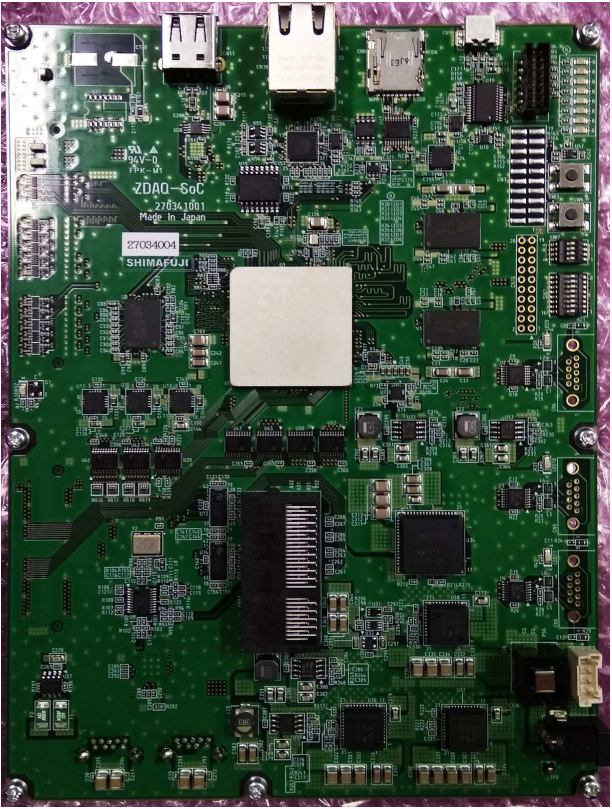}
\end{center}
\caption{Photo of the ZDAQ board.  Zynq SoC chip is placed at the center of the board. \label{fig:photo}}
\end{figure}

Following the concept outlined in the previous section, we have developed the data acquisition board named ZDAQ with a Zynq SoC chip. The board development was performed in collaboration with a group comprised of space-based astronomy, ground-based observatory, and medical instruments, for application within their respective fields. Shimafuji Electric Inc. designed the circuit and developed the board. The photo of the ZDAQ board is shown Fig.~\ref{fig:photo}.
 The size of the board is 200 $\times$ 150 mm$^2$. We used a Zynq chip XC7Z045-1FFG900 to ensure sufficient input/output (I/O) pins to control and read out the CMOS sensor. ZDAQ has I/O connectors for 48 channels of Low Voltage Differential Signal (LVDS) I/O signals and 34 channels input and 66 channel output 3.3 V Low Voltage CMOS (LVCMOS33) signals. The PS of the ZDAQ Zynq device is a dual core processor Advanced RISC Machine (ARM) Cortex-A9 \cite{cortex}, and the clock frequency for the processor is 667 MHz. The PS memory is 1 GB Double-Data-Rate 3 (DDR 3) Synchronous Dynamic Random Access Memory (SDRAM), and 256 MB DDR 2 SDRAM is also mounted for buffering the image data from the CMOS sensor and other applications. A Universal Serial Bus (USB) 3.0 port and four lanes of Peripheral Component Interconnect Express 2.0 (PCI Express Gen2) are mounted to connect the data storage device. Data transfer speeds of the USB 3.0 and four lanes PCI Express Gen2 are 5 and 16 Gbps, respectively, sufficient for the requirement of the CMOS sensor readout system for the sounding rocket observation. ZDAQ also has three space wire ports and a Gigabit Ethernet port for multiple applications. The specifications of the ZDAQ board are summarized in Table~\ref{table:spec}. In the sounding rocket experiment, the radiation tolerance requirement is not strong owing to the short flight. It is necessary for the system to pass radiation tolerance tests for satellite applications.
\begin{table}[ht!]
\centering
\caption{Specification of the ZDAQ-SoC board.} \label{table:spec}
\begin{tabular}{ll}
\hline
Component & Description \\
\hline
SoC & ZYNQ-7000 SoC (XC7Z045-1FFG900)\\
Processor core & ARM Cortex A9 DualCore with NEON \\
 & Clock: 667 MHz\\
 & Command / data cache: 32 kB each\\
 & L2 cache: 512 kB\\
System memory & DDR3-SDRAM 1 GB\\
Data buffer & DDR2-SDRAM 256 MB\\
USB& USB2.0 host $\times$1, USB3.0 host $\times$2\\
PCI Express & Gen2 $\times$ 4\\
Ethernet &  1000Base $\times$ 1\\
SpaceWire &  $\times$ 3\\
LVDS I/O & 48 ch\\
LVCMOS33 I/O & Input: 34 ch, Output: 66 ch\\
Power & 0.53 A typical for 12 V\\
Board size & 200 $\times$ 150 mm\\
\hline
\end{tabular}
\end{table}

We can utilize the ZDAQ board by developing a board to stack on the ZDAQ board. For example, we have developed a board to interface the ZDAQ with the CMOS sensor. The interface board will be stacked on ZDAQ for the rocket flight, and I/O connectors are then directly connected to the ZDAQ board.

\begin{figure}[ht!]
\begin{center}
\includegraphics[width=6.5cm]{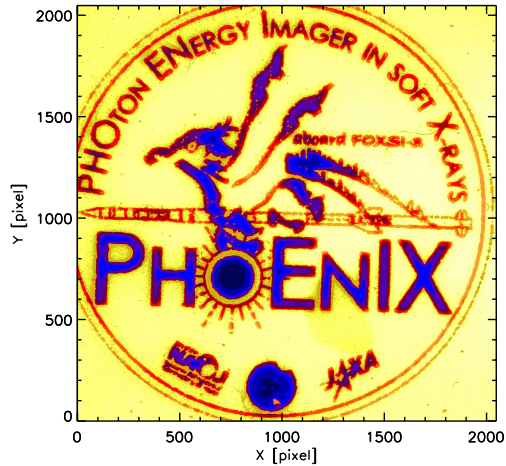}
\end{center}
\caption{Example of an image taken by the system, with visible light masked by the PhoEnIX logo printed on an overhead projector film.  \label{fig:image}}
\end{figure}

Using the ZDAQ board, we successfully controlled the CMOS sensor for the rocket flight and recorded an image. An example of the image taken in visible light is illustrated in Fig.~\ref{fig:image}. The control signals are generated in PL, and CMOS sensor operation can be controlled by registers from PS. We transferred the image data from the CMOS sensor received by the PL to a reserved memory area in the DDR3 SDRAM using the Advanced eXtensible Interface 4 (AXI4) protocol. The AXI4 bus has a throughput of up to 600 MB/s, which sufficient for the image data transfer. We also successfully recorded the data to storage devices by running Linux in PS. We tested a Solid State Drive (SSD) connected by USB 3.0, and a non-volatile memory express (NVMe) \cite{nvme} SSD connected by the PCI Express port. Using a computer with a high-speed CPU, USB 3.0 SSDs show the sequential read and write speed of $>$300 MB/s, and NVMe SSDs demonstrated speeds $>$1000 MB/s. We measured the end-to-end throughput from the CMOS sensor to the data storage using a Samsung 960 EVO NVMe SSD device. The performance was significantly improved with the direct I/O operation in Linux \cite{joyce2016}, and the data transfer speed was approximately 160 MB/s. The difference between the speed using the fast computer and that using our system is attributed to the difference in the clock frequencies of the processors. We thus confirmed that the system has sufficient speed for the high-speed X-ray imaging spectroscopy using the CMOS sensor. We used a SSD with 480 GB volume, large enough for the rocket requirement. The average power of the entire system, including the CMOS sensor and SSD during continuous CMOS sensor operation and data record, was measured to be approximately 11 W. A battery for the sounding rocket flight is able to provide this amount of power.

We have developed the ZDAQ board, not only to specify the CMOS readout system, but also for high-speed data transfer and/or large volume data recording. Many applications are possible with stacking a board on ZDAQ. Another example that we have developed is a general-purpose analog electronics board named ZDAQ-ANALOG-1. ZDAQ-ANALOG-1 has 200 Msps 12 bit analog-to-digital convertors, digital-to-analog convertors, a direct digital synthesizer, and an arbitrary waveform generator. It can be easily connected to the ZDAQ board, and we can control their functions through an operating system running on the Zynq PS.

\section{Summary}
We have developed the X-ray detector system with a high-speed back-illuminated CMOS sensor for soft X-ray imaging and spectroscopy of the Sun. To record high-speed data during the sounding rocket experiment, we have developed the data acquisition board, ZDAQ, with the Zynq SoC. PL controlled the sensor with synchronous signals, and PS controlled the data recording with the flexibility of software. We successfully tested the functions to control the CMOS sensor and to write data to the storage device. We expect to perform the first photon-counting soft X-ray imaging spectroscopy of the Sun utilizing this system in the FOXSI-3 sounding rocket experiment, in the summer of 2018.

\section*{Acknowledgment} 
We thank Kento Furukawa for his tactful assistance for taking the example image. 
This work was supported by JSPS KAKENHI Grant Numbers 17H04832, 16H02170, 15H03647, 15H05892 and 24105007.





\end{document}